\documentclass[a4,12pt]{article}

\usepackage{amsmath,amssymb,bm}
\usepackage{mathrsfs}
\usepackage{graphicx,color}
\usepackage{cite}
\usepackage{cases}

\def\tfontsize{scaled\magstep4}
\font\titlerm=cmr10 \tfontsize

\makeatletter

\@addtoreset{equation}{section}

\renewcommand{\section}{\@startsection{section}{1}{\z@}
	{-3.5ex \@plus -1ex \@minus -.2ex}{2.3ex \@plus.2ex}
	{\normalfont\normalsize\bfseries}}
\renewcommand{\subsection}{\@startsection{subsection}{2}{\z@}
	{-3.5ex \@plus -1ex \@minus -.2ex}{2.3ex \@plus.2ex}
	{\normalfont\normalsize\bfseries}}
\renewcommand{\subsubsection}{\@startsection{subsubsection}{3}{\z@}
	{-3.5ex \@plus -1ex \@minus -.2ex}{2.3ex \@plus.2ex}
	{\normalfont\normalsize\it}}
\makeatother

\newcommand\tr{\mathop{\rm tr}\nolimits}
\renewcommand\Im{\mathop{\rm Im}\nolimits}

\newcommand\rrangle{\rangle\!\rangle}
\newcommand\llangle{\langle\!\langle}

\newcommand\bra[2]{{}_{#2}\!\langle #1}
\newcommand\ket[2]{#1 \rangle\!_{#2}}
\newcommand\bbra[2]{{}_{#2}\!\llangle #1}
\newcommand\kket[2]{#1 \rrangle\!_{#2}}

\newcommand\vk{{\bm k}}
\newcommand\vl{{\bm l}}
\newcommand\vp{{\bm p}}
\newcommand\vq{{\bm q}}
\newcommand\vr{{\bm r}}

\newcommand\CA{{\cal A}}
\newcommand\CH{{\cal H}}
\newcommand\CN{{\cal N}}
\newcommand\CO{{\cal O}}
\newcommand\CP{{\cal P}}
\newcommand\CS{{\cal S}}
\newcommand\CT{{\cal T}}

\newcommand\bal{\begin{align}}
\newcommand\eal{\end{align}}

\begin{document}
\renewcommand{\thefootnote}{\fnsymbol{footnote}}

\begin{titlepage}
\begin{flushright}
\end{flushright}
\setlength{\baselineskip}{19pt}
\bigskip\bigskip\bigskip

\vbox{\centerline{\titlerm Entanglement in Elastic and Inelastic}
\bigskip
\centerline{\titlerm Two-particle Scatterings at High Energy}
}

\bigskip\bigskip\bigskip

\centerline{Robi Peschanski\footnote{\tt robi.peschanski@ipht.fr}}
\medskip
{\it 
\centerline{ Institut de Physique Th{\' e}orique\footnote{Unit\'e Mixte de Recherche 3681 du CNRS}, Universit\'e Paris-Saclay, CEA}
\centerline{F-91191 Gif-sur-Yvette, France}
}
\bigskip
\centerline{and}
\bigskip
\centerline{Shigenori Seki\footnote{\tt shigenori.seki@setsunan.ac.jp}}
\medskip
{\it 
\centerline{Institute for Fundamental Sciences, Setsunan University}
\centerline{17-8 Ikedanaka-machi, Neyagawa, Osaka 572-8508, Japan}
}

\vskip .3in

\centerline{\bf Abstract}

We study the entanglement produced in transverse momentum by two-particle scattering at high energy. 
Employing the S-matrix framework for the derivation of reduced density matrices, 
we formulate the entanglement entropy for an inelastic scattering 
as well as an elastic one. 
We display the formulas of the entanglement entropy in terms of two-body cross sections.
We also derive the entanglement density as a function of the transverse momentum.  
As an application, we then focus on both forward  elastic ($pn \to pn$) and  inelastic 
($pn \to np$) channels  scattering allowing for a fruitful comparison of the two reactions 
with the same proton-neutron content. We evaluate the elastic and inelastic entanglement 
entropy by using known parameterizations of experimental data for neutron-proton reactions. 
Comparing those entanglement entropies, we observe that the inelastic scattering produces 
more overall entanglement than the elastic one in the $pn$ sector.

\vfill\noindent
30 January 2026
\end{titlepage}

\renewcommand{\thefootnote}{\arabic{footnote}}
\setcounter{footnote}{0}
\setlength{\baselineskip}{19pt}

\section{Introduction}

Quantum entanglement is one of the most significant phenomena in quantum physics, 
and many scientists in various fields have been studying it. 
In order to measure the strength of entanglement, one can use entanglement entropy. 
The density matrix, $\rho$, of a two-body state 
in a Hilbert space, $\CH_A \otimes \CH_B$, leads to the reduced density matrix, 
$\rho_A = \tr_B \rho$. 
The entanglement entropy of that state is defined by $-\tr_A \rho_A \ln \rho_A$.
We are interested in the entanglement of two-body scattering. 
Even if the initial state of incoming particles is not entangled, 
the final state of outgoing particles should be entangled by interactions. 
So we ask a simple question, 
how much entanglement is produced by a two-body scattering process.

We consider the entanglement of the final state after scattering. 
In general, one can consider entanglements on various quantum numbers. 
For example, an entanglement of spin is one of the most well-known entanglements. 
We would like to reveal the entanglement on a momentum Hilbert space. 
The momentum-space entanglement in quantum field theory has been considered by Ref.~\cite{BMR}.
In Ref.~\cite{PS1}, the entanglement entropy of the final state in an elastic scattering 
has been formulated by using an S-matrix theory. 
The formula of entanglement entropy in Ref.~\cite{PS1} is described in terms of physical quantities, 
{\it i.e.}, cross sections. 
However it includes a divergent factor yielded by the infinite volume of the momentum Hilbert space. 
In order to avoid this divergence, Ref.~\cite{PS2} has suggested the volume regularization, 
and has concretely evaluated such regularized entanglement entropy for proton-proton elastic scattering 
with parameterizations of the experimental data given by Tevatron in Fermilab and Large Hadron Collider in CERN. 
We can also see many other articles on the entanglement related to particle scattering, 
for instance, on 
scattering in QED \cite{ToTo,RNMM,FDR,FHAPS,FS,FMBS,BDLMM1,LLY,BDLMM2}, 
fermion-fermion scattering \cite{FDH,FL}, 
deep inelastic scattering \cite{KL,RM1,Le}, 
and so on\cite{GS,BHSST,RM2,Mi,GKP,BLM,AERS,KS,LY,ThTr,LTWZZ,SWZ}. 

In this paper, we consider the entanglement entropy of final states 
in inelastic scatterings as well as an elastic scattering. 
Although more than two particles can appear, and many more at high energies, as a final state in actual scattering processes, 
we concentrate on the entanglement of the two-particle final-state channels. 
We shall formulate the entanglement entropy of the two-particle final state for inelastic scatterings 
by using the S-matrix theory, as Refs.~\cite{PS1,PS2} have done for an elastic scattering, keeping both derivations in parallel.

We shall concretely evaluate the entanglement entropy in elastic and inelastic channels. 
Although we can consider many examples of scattering, 
we focus on a proton-neutron scattering, which allows for  elastic or inelastic channels, depending on considering the forward region in $pn \to pn$ or $pn \to np$ scattering. 
There are thus two channels; $pn \to pn$ and $pn \to np$, with same initial and final particle content. We call the former the elastic channel 
and the latter the inelastic one. We distinguish them in our formulation 
from the viewpoint of the exchange of $t$-channel quantum numbers. 
We can practically find the appropriate parameterizations of experimental data for both channels from various reactions of neutron projectiles on fixed target experiments at high energy.  

As an interesting by-product of our analysis, we obtain the density distribution of the entanglement entropy as a function of transverse momentum, allowing for obtaining a hint of the entanglement flow during the scattering process in both elastic and inelastic cases.

In Section 2, we review the S-matrix formalism with partial wave expansions \cite{Va,BV}. 
In Section 3, using this formalism, we define the density matrices and formulate the entanglement entropy 
for two-particle final states both in elastic and inelastic channels. We then obtain the distribution density of the entanglement entropy as a function of momentum transfer.
In Section 4, we evaluate the entanglement entropy and its transverse momentum distribution for  proton-neutron scattering from known parameterizations of the experimental data, and clarify the difference of entanglement entropy between the elastic and inelastic channels. 
Section 5 is dedicated to  discussion, conclusions and outlook.

\section{Entanglement entropy in the S-matrix formulation}

Since we are interested in the entanglement entropy of general two-particle states at high energy, 
we are led to consider three kinds of scattering channels as follows:
\begin{align}
A_1 B_1 \to 
\begin{cases}
A_1 B_1 & \text{(elastic channel)} \\
A_n B_n \quad (n=2,3,\dots) & \text{(two-particle inelastic channels)} \\
X & \text{(more than two-particle channels)}
\end{cases} \label{scchannels}
\end{align}
The initial state consists of two particles, $A_1$ and $B_1$, 
and the final states are $A_1 B_1$, $A_n B_n$ $(n=2,3,\dots)$ and the multi-particle  states, $X$.
In the two-particle inelastic channels, 
the pair of $A_1$ and $B_1$ is different from the one of $A_2$ and $B_2$, 
but it is not necessary that both $A_1$ and $B_1$ are different from $A_n$ and $B_n$,
namely, for example,  combinations of $A_n = A_1$ and $B_n \neq B_1$ are allowed. 
Furthermore, $A_1B_1$ may be different from $B_1A_1$ in our formulation and classified into the inelastic channels, if corresponding to a scattering with non-vacuum quantum number exchange. 

We shall formulate the entanglement entropy of two-particle states in terms of the S-matrix formalism.
For sake of simplicity we will stick to the formalism with only one inelastic channel, {\it i.e.}, 
\begin{align}
A_1 B_1 \to 
\begin{cases}
A_1 B_1 & \text{(elastic channel)} \\
A_2 B_2 & \text{(two-particle inelastic channel)} \\
X & \text{(more than two-particle channels)}
\end{cases} \label{scchannel2}
\end{align}
at high energy \cite{Va} but the generalization of the formalism to a finite set of two-body inelastic channels has also been done in  matrix form \cite{BV}, which can be easily applied.

\subsection{S-matrix and unitarity}

Following Ref.~\cite{Va}, we shall adjust the S-matrix theory for the scattering 
including a two-particle inelastic channel together with the elastic one. 

The Hilbert spaces of two particle states, $A_1 B_1$ and $A_2 B_2$, are respectively
\begin{align}
\CH_1 = \CH_{A_1} \otimes \CH_{B_1}\,, \quad \CH_2 = \CH_{A_2} \otimes \CH_{B_2} \,.
\end{align}
Each Hilbert space $\CH_I$ ($I=A_1,B_1,A_2,B_2$) is the Fock space of the particle $I$.  
The state of $I$ with momentum $\vp$ in $\CH_I$ is denoted by $|\ket{\vp}{I} \in \CH_I$.
The inner products of the states is defined as 
\begin{align}
\bra{\vp}{I}|\ket{\vq}{J} = \delta_{IJ}\sqrt{2E_{I\vp}\,2E_{J\vq}} \delta^{(3)}(\vp-\vq) \,, 
\end{align}
where the energy $E_{I\vp} = \sqrt{{m_I}^2 +p^2}$ $(p\equiv|\vp|)$.
The total Hilbert space concerning the scattering process \eqref{scchannel2} is given by 
$\CH_{\rm tot} = \CH_1 \otimes \CH_2 \otimes \CH_X$, 
where $\CH_X$ is the Hilbert space for multi-particle states, {\it i.e.}~for more than two particles in the final state.
We describe states in $\CH_1$ and $\CH_2$ as 
\begin{align}
|\ket{\vp,\vq}{1} \equiv |\ket{\vp}{A_1} \otimes |\ket{\vq}{B_1} \in \CH_1 \,, \quad 
|\ket{\vp,\vq}{2} \equiv |\ket{\vp}{A_2} \otimes |\ket{\vq}{B_2} \in \CH_2 \,, \label{ABandCDstates}
\end{align}
for the sake of convenience. 
The complete set given by orthogonal basis satisfies 
\begin{align}
{\bf 1} &= 
	\int {d^3\vp d^3\vq \over 2E_{A_1\vp} 2E_{B_1\vq}} |\ket{\vp,\vq}{1}\, \bra{\vp,\vq}{1}|
	+\int {d^3\vp d^3\vq \over 2E_{A_2\vp} 2E_{B_2\vq}} |\ket{\vp,\vq}{2}\, \bra{\vp,\vq}{2}|
	+\int dX\, |X\rangle \langle X| \,, \label{cmpset}
\end{align}
where ${\bf 1}$ is the unit matrix.

We introduce the well-known S-matrix, $\CS$, relating the set of asymptotic final states to the initial pair of states. The transition T-matrix, $\CT$, which describes the scattering amplitudes is such that 
$\CS = {\bf 1} + 2i\CT$. 
The unitarity of the S-matrix corresponding to the completeness of the final state basis writes $\CS^\dagger \CS = {\bf 1}$. 
It is equivalently expressed in terms of the T-matrix, 
\begin{align}
{1 \over 2i}(\CT - \CT^\dagger) = \CT^\dagger \CT \,. \label{Tunitarity}
\end{align}
In terms of Eq.~\eqref{cmpset}, the unitarity condition \eqref{Tunitarity} leads to 
a relation of the T-matrix elements concerning the two-particle states, 
$\bra{\vp,\vq}{i}|$ and $|\ket{\vp',\vq'}{j}$;
\begin{align}
&{1 \over 2i} \bigl\{ \bra{\vp,\vq}{i}|\CT | \ket{\vp',\vq'}{j} - \bra{\vp,\vq}{i}|\CT^\dagger | \ket{\vp',\vq'}{j} \bigr\} \nonumber \\
&= \sum_{h=1}^2 \int {d^3\vp''\over 2E_{A_h\vp''}}{d^3\vq'' \over 2E_{B_h\vq''}}\, \bra{\vp,\vq}{i}|\CT^\dagger|\ket{\vp'',\vq''}{h}\, \bra{\vp'',\vq''}{h}| \CT | \ket{\vp',\vq'}{j} \nonumber \\
&\quad +\int dX\, \bra{\vp,\vq}{i}|\CT^\dagger|X\rangle \langle X|\CT | \ket{\vp',\vq'}{j}  \quad (i,j = 1,2)\,. \label{eq:unitarity}
\end{align}

\subsection{Overlap matrix and partial wave expansion}

Let us adopt the center-of-mass frame.  
We describe the states \eqref{ABandCDstates} as  
\begin{align}
|\kket{\vp}{1} \equiv |\ket{\vp,-\vp}{1} \,, \quad 
|\kket{\vp}{2} \equiv |\ket{\vp,-\vp}{2} \,,
\end{align}
and the inner product of states becomes 
\begin{align}
&\bbra{\vp}{i}|\kket{\vq}{j} = \delta_{ij}2E_{A_i\vp}\,2E_{B_i\vp} \delta^{(3)}(\vp-\vq) \delta^{(3)}(0) \quad (i,j =1,2) \,.
\end{align}
In terms of these expressions, the complete set of orthogonal basis \eqref{cmpset} is 
rewritten as 
\begin{align}
{\bf 1} = \sum_{h=1}^2\int {d^3\vp \over 2E_{A_h\vp}\,2E_{B_h\vp} \delta^{(3)}(0)} |\kket{\vp}{h}\, \bbra{\vp}{h}|
+\int dX\, |X\rangle \langle X| \,. \label{completeness}
\end{align}

We extract the factors of energy-momentum conservation from the T-matrix elements, 
\begin{align}
\bbra{\vp}{i} | \CT | \kket{\vq}{j} = \bbra{\vp}{i} | {\bf t} |\kket{\vq}{j} \,\delta^{(4)}(\mathscr{P}_{i\vp} -\mathscr{P}_{j\vq}) \,, \quad 
\bbra{\vp}{i} | \CT |X\rangle = \bbra{\vp}{i} | {\bf t}|X\rangle \delta^{(4)}(\mathscr{P}_{i\vp} -\mathscr{P}_X) \,, \label{extractconv}
\end{align}
where $\mathscr{P}_{i\vp}$ and $\mathscr{P}_X$ are the total energy-momenta 
of the states $|\vp\rrangle_i$ and $|X\rangle$ respectively.  
For example, the first delta function in \eqref{extractconv} stands for
\begin{align}
\delta^{(4)}(\mathscr{P}_{i\vp} -\mathscr{P}_{j\vq}) 
&= \delta^{(3)}((\vp+(-\vp))-(\vq+(-\vq)))\ \delta((E_{A_i\vp}+E_{B_i\vp})-(E_{A_j\vq}+E_{B_j\vq})) \nonumber\\
&= \delta^{(3)}(0)\ \delta((E_{A_i\vp}+E_{B_i\vp})-(E_{A_j\vq}+E_{B_j\vq})) \,.
\end{align}
Since some factors of energy-momentum conservation in Eq.~\eqref{eq:unitarity} are cancelled 
in the center-of-mass frame, 
one can rewrite the condition \eqref{eq:unitarity} as 
\begin{align}
& {1 \over 2i}\bigl(\bbra{\vq}{i}|{\bf t} |\kket{\vk}{j} 
	-\bbra{\vq}{i} | {\bf t}^\dagger |\kket{\vk}{j} \bigr) \nonumber\\
&= \sum_{h=1}^2 \int{d^3\vq \over 2E_{A_h\vq}\,2E_{B_h\vq} \delta^{(3)}(0)} 
	\bbra{\vq}{i} |{\bf t}^\dagger | \kket{\vq}{h}\, 
	\delta^{(4)}(\mathscr{P}_{h\vq} -\mathscr{P}_{j\vk})\, 
	\bbra{\vq}{h} |{\bf t}|\kket{\vk}{j} 
	+F_{ij}(\vp,\vk) \,, \label{unicondTe}
\end{align}
where $F_{ij}(\vp,\vk)$ is an overlap matrix \cite{Va}, 
\begin{align}
F_{ij}(\vp,\vk) = \int dX\,  \bbra{\vp}{i} |{\bf t}^\dagger |X\rangle \delta^{(4)}(P_{X} -P_{j\vk}) \langle X|{\bf t}|\kket{\vk}{j} \,. \label{OLM}
\end{align}
The delta functions for energy conservation in Eq.~\eqref{unicondTe} imply that 
the center-of-mass energies of the states are common, that is, 
\begin{align}
E_{A_i\vp} + E_{B_i\vp} = E_{A_j\vk} + E_{B_j\vk} = E_{A_h\vq} + E_{B_h\vq} \equiv \sqrt{s} \,. \
\end{align}
$s$ is the usual Mandelstam variable. 

We expand a T-matrix element and the overlap matrix in partial waves, 
\begin{align}
\bbra{\vp}{i}|{\bf t}|\kket{\vk}{j} 
&= {\sqrt{s} \over \pi\sqrt{pk}}\sum_{\ell=1}^\infty (2\ell+1)\tau_{ij}^{\ell} P_\ell(\cos\theta) \,, \label{tmatPWE} \\
F_{ij}(\vp,\vk) 
&= {\sqrt{s} \over 2\pi\sqrt{pk}} \sum_{\ell=1}^\infty (2\ell+1)f_{ij}^{\ell} P_\ell(\cos\theta) \,, \label{overlapPWE}
\end{align}
where the scattering angle, $\theta$, is defined by $\vp\cdot\vk = pk\cos\theta$. 
Then, by using partial wave modes in Eqs.~\eqref{tmatPWE} and \eqref{overlapPWE}, 
one can express the unitarity condition \eqref{unicondTe} in the following simple form; 
\begin{align}
\Im \tau_{ij}^\ell = \sum_{h=1}^2 \tau_{ih}^{\ell*} \tau_{hj}^\ell + {1 \over 2} f_{ij}^\ell \,. \label{PWunicond}
\end{align}

\subsection{Cross sections} \label{subsec:cs}

The two-body scattering amplitude $\CA_{ij}(s,t)$ is of the channel: $A_j B_j \to A_i B_i$ 
with the initial state, $|\kket{\vk}{j}$, and the final state, $|\kket{\vp}{i}$. 
It is related with the S-matrix element, $\bbra{\vp}{i}|{\bf s}|\kket{\vk}{j}$, as follows:
\begin{align}
\bbra{\vp}{i}|{\bf s}|\kket{\vk}{j} = {\sqrt{s} \over \pi\sqrt{pk}}\cdot 
2\biggl(\delta(1-\cos\theta)\delta_{ij} +{i \over 16\pi}\CA_{ij}(s,t)\biggr) \,.
\end{align}
${\bf s}$ is defined by ${\bf s} = {\bf 1} + 2i{\bf t}$, while
$s$ and $t$ are the Mandelstam variables, here  given by 
\begin{align}
\sqrt{s}=\sqrt{m_{A_j}^2+k^2} +\sqrt{m_{B_j}^2+k^2}\ ,\quad t=2k^2(\cos\theta -1) \,.
 \label{Mandelstam}
\end{align}

Thus $\CA_{ij}$ is described in terms of the partial wave modes as 
\begin{align}
\CA_{ij} = 16\pi\sum_{\ell=0}^\infty (2\ell+1)\tau_{ij}^\ell P_\ell(\cos\theta) \,.
\end{align}
Since the differential cross section is 
\begin{align}
{d\sigma_{ij} \over dt} = {\pi \over k^4}\biggl|\sum_{\ell=0}^\infty (2\ell+1)\tau_{ij}^\ell P_\ell(\cos\theta)\biggr|^2 
= {|\CA_{ij}|^2 \over 256\pi k^4} \,, \label{diffcross}
\end{align}
we obtain the cross section of the two-particle final states, thanks to the orthogonality property of Legendre polynomials, 
\begin{align}
\sigma_{ij} = {4\pi \over k^2} \sum_{\ell=0}^\infty (2\ell+1) |\tau_{ij}^\ell|^2 \,. \label{ijcross}
\end{align}
If $i=j$, $\sigma_{ii}$ is the elastic cross section,  
while if $i\neq j$, $\sigma_{ij}$ is the two-particle inelastic cross section. 
One can also obtain the cross section for multi-particle inelastic channel from the overlap matrix modes, namely 
\begin{align}
\sigma_{ij}^X = {2\pi \over k^2} \sum_{\ell=0}^\infty (2\ell+1) f_{ij}^\ell \,. \label{Xcross}
\end{align}

When we set the initial state to be $A_1 B_1$, the unitarity condition \eqref{PWunicond} 
implies 
\begin{align}
\Im \tau_{11}^\ell = |\tau_{11}^\ell|^2 + |\tau_{21}^\ell|^2 + {1 \over 2} f_{11}^\ell \,.
\end{align} 
Then the total cross section is given by 
\begin{align}
\sigma_{11}^{\rm tot} = {4\pi \over k^2} \sum_{\ell=0}^\infty (2\ell+1) \Im \tau_{11}^\ell
= \sigma_{11}+\sigma_{21}+\sigma_{11}^X \,. \label{totcross}
\end{align}

\section{Formulation of entanglement entropy}

\subsection{Density matrices of two-body elastic and inelastic final states}

We consider the scattering processes \eqref{scchannel2}. 
The two particles, $A_1 B_1$, in the initial state $|{\rm ini}\rangle$ have momenta, 
$\vk$ and $\vl$, respectively, {\it i.e.},
$|{\rm ini}\rangle = |\vk,\vl\rangle_1$. 
Then, in terms of the S-matrix, the final state is given by $\CS|{\rm ini}\rangle \in \CH_{\rm tot}$. 
By using the projection from $\CH_{\rm tot}$ to $\CH_1$, 
we obtain the final state as $A_1 B_1$ in the elastic channel, 
\begin{align}
|\ket{{\rm fin}}{1} 
= \left( \int {d^3\vp \over 2E_{A_1\vp}}{d^3\vq \over 2E_{B_1\vq}}\, |\ket{\vp,\vq}{1}\, \bra{\vp,\vq}{1}|\right) \CS | \ket{\vk,\vl}{1} 
&=\!\! \int \!{d^3\vp \over 2E_{A_1\vp}}{d^3\vq \over 2E_{B_1\vq}}\, |\ket{\vp,\vq}{1}\, \bra{\vp,\vq}{1}| (1\!+\!2i\CT) | \ket{\vk,\vl}{1} \,. \label{finstel}
\end{align}
In the same way, the final state as $A_2 B_2$ in the two-particle inelastic channel is 
\begin{align}
\!\!\!\!\!\!\!|\ket{{\rm fin}}{2}
&= \left( \int {d^3\vp \over 2E_{A_2\vp}}{d^3\vq \over 2E_{B_2\vq}}\, |\ket{\vp,\vq}{2}\, \bra{\vp,\vq}{2}|\right) \CS | \ket{\vk,\vl}{1} 
= \int {d^3\vp \over 2E_{A_2\vp}}{d^3\vq \over 2E_{B_2\vq}}\, |\ket{\vp,\vq}{2}\, \bra{\vp,\vq}{2}| 2i\CT | \ket{\vk,\vl}{1} \,. \label{finstinel}
\end{align}
Note the essential difference between Eqs.~\eqref{finstel} and  \eqref{finstinel}: The unit matrix in $\CS$ does not contribute to the final state in the inelastic channel 
due to $\bra{\vp,\vq}{2}| {\bf 1} | \ket{\vk,\vl}{1} = 0$. 

By using these final states, we define the total density matrices of final two-body states as 
\begin{align}
\rho_1 = {1 \over \CN_1} |\ket{{\rm fin}}{1}\, \bra{{\rm fin}}{1}| \,, \quad 
\rho_2 = {1 \over \CN_2} |\ket{{\rm fin}}{2}\, \bra{{\rm fin}}{2}| \,.
\end{align}
$\CN_i$ ($i=1,2$) are normalization factors which are determined by $\tr_{A_i}\tr_{B_i}\rho_i = 1$. 

We then obtain reduced density matrices from the total density matrices 
by taking trace on one of the Hilbert spaces of two particles. 
In the case of elastic channel, we calculate the reduced density matrix, $\rho_{A_1} = \tr_{B_1} \rho_1$, 
with extracting the factor of energy-momentum conservation from the expression \eqref{finstel}, 
\begin{align}
\rho_{A_1} &= \int {d^3\vr \over 2E_{B_1\vr}}\, \bra{\vr}{B_1}| \rho_1 |\ket{\vr}{B_1} \nonumber\\
&= {1 \over \CN_1}\int {d^3\vp \over 2E_{A_1\vp}}\,{\delta(E_{A_1\vp}\!+\!E_{B_1(\vk+\vl-\vp)}\!-\!E_{A_1\vk}\!-\!E_{B_1\vl}) \delta(0) \over 2E_{A_1\vp}\,2E_{B_1(\vk+\vl-\vp)}} 
|\bra{\vp,\vk+\vl-\vp}{1}|{\bf s}|\ket{\vk,\vl}{1}|^2\,
|\ket{\vp}{A_1}\,\bra{\vp}{A_1}| \,. \label{ELredDMref}
\end{align}
Although there is a divergent factor, $\delta(0)$, we leave it for the present 
and shall regularize it later. 
In the same way, the reduced density matrix, $\rho_{A_2} = \tr_{B_2} \rho_2$, for the case of two-particle inelastic channel is described as 
\begin{align}
\rho_{A_2} &= \int {d^3\vr \over 2E_{B_2\vr}}\, \bra{\vr}{B_2}| \rho_2 |\ket{\vr}{B_2} \nonumber\\
&= {1 \over \CN_2}\int {d^3\vp \over 2E_{A_2\vp}}\,{\delta(E_{A_2\vp}\!+\!E_{B_2(\vk+\vl-\vp)}\!-\!E_{A_1\vk}\!-\!E_{B_1\vl}) \delta(0) \over 2E_{A_2\vp}\,2E_{B_2(\vk\!+\!\vl\!-\!\vp)}} 
\, |\bra{\vp,\!\vk\!+\!\vl\!-\!\vp}{2}|{2{\bf t}}|\ket{\vk,\vl}{1}|^2
|\ket{\vp}{A_2}\,\bra{\vp}{A_2}| \,. \label{INELredDMref}
\end{align}
Note again the difference between the matrix elements in Eqs.~\eqref{ELredDMref} and \eqref{INELredDMref}.

\subsection{Entanglement entropy}

In this subsection, we adopt the center-of-mass frame again. 
The momenta of the initial state satisfy $\vl = -\vk$, 
and the center-of-mass energy is $\sqrt{s} = E_{A_1\vk}+E_{B_1\vk}$. 

We firstly calculate $\tr_{A_i} (\rho_{A_i})^n$, which is related to 
the R{\' e}nyi entropy, $(1-n)^{-1} \ln \tr_{A_i}(\rho_{A_i})^n$. 
Then it gives us the entanglement entropy through 
\begin{align}
S_{i1} = -\tr_{A_i}\rho_{A_i}\ln \rho_{A_i}
= -\lim_{n\to 1}{\partial \over \partial n} \tr_{A_i}(\rho_{A_i})^n \,. \label{renyiEE}
\end{align}
$S_{i1}$ denotes the entanglement entropy of the final state in the channel: $A_1 B_1 \to A_i B_i$. 
This observable serves as a measure of the entanglement between the two particles in each final state.

\subsubsection{Elastic channel}

Since the entanglement entropy between two particles in the elastic scattering has been 
formulated by Ref.~\cite{PS1}, 
we here calculate the elastic entanglement entropy in our current notation  allowing for 
a parallel calculation between the elastic and inelastic cases.

In the center-of-mass frame, the reduced density matrix \eqref{ELredDMref} becomes 
\begin{align}
\rho_{A_1} &= {1 \over \CN_1} \int {d^3\vp \over 2E_{A_1\vp}}\,
{\delta(p-k) \delta(0) \over 4p\sqrt{s}} 
|\bbra{\vp}{1}|{\bf s}|\kket{\vk}{1}|^2\,
|\ket{\vp}{A_1}\,\bra{\vp}{A_1}| \,,
\end{align}
where we used 
\begin{align}
\delta(E_{A_1\vp}+E_{B_1\vp}-\sqrt{s}) = {2E_{A_1\vp}\,2E_{B_1\vp} \over 4p\sqrt{s}}\ \delta(p-k) \,.
\end{align}
The normalization factor $\CN_1$ is determined by $1 = \tr_{A_1}\tr_{B_1}\rho_1 = \tr_{A_1} \rho_{A_1}$, 
so that 
\begin{align}
\CN_1 = \delta(0) \delta^{(3)}(0) \int d^3\vp\, {\delta(p-k) \over 4p\sqrt{s}} 
|\bbra{\vp}{1}|{\bf s}|\kket{\vk}{1}|^2 \,. \label{nor1}
\end{align}
Then we calculate 
\begin{align}
\tr_{A_1}(\rho_{A_1})^n 
= \int d^3\vp\, \delta^{(3)}(0) \biggl( {\delta(p-k) \delta(0) \over \CN_1\cdot 4p\sqrt{s}} 
|\bbra{\vp}{1}|{\bf s}|\kket{\vk}{1}|^2 \biggr)^n \,. \label{eltrrhon}
\end{align}

Now we use the partial wave expansion \eqref{tmatPWE} and \eqref{overlapPWE}. 
The normalization factor \eqref{nor1} is written as 
\begin{align}
\CN_1 = \delta(0) \delta^{(3)}(0) {\sqrt{s} \over \pi k}\sum_{\ell=0}^\infty (2\ell+1)|s_{11}^\ell|^2 \,,
\end{align}
where the partial wave mode $s_{ij}^\ell$ is defined as 
$s_{ij}^\ell = \delta_{ij} +2i\tau_{ij}^\ell$. 
Then Eq.~\eqref{eltrrhon} becomes 
\begin{align}
\tr_{A_1}(\rho_{A_1})^n 
&= \int_{-1}^1 d\!\cos\theta\, \biggl({\delta(0) \over 2\pi k^2 \delta^{(3)}(0)}\biggr)^{n-1} 
(\CP_{11}(\cos\theta))^n \,, \label{trA1rhon2} \\ 
\CP_{11}(\cos\theta) &\equiv {|\sum_{\ell=0}^\infty (2\ell+1)s_{11}^\ell P_\ell(\cos\theta)|^2 \over 2 \sum_{\ell=0}^\infty (2\ell+1)|s_{11}^\ell |^2} \,, \label{elp1}
\end{align}
after the integration over the spherical coordinates of $\vp$ with azimuthal symmetry. 

The ``volume'' $V$ of momentum Hilbert space is infinite. 
Indeed it can be formally written  as 
\begin{align}
V \equiv \sum_{\ell=0}^\infty (2\ell+1) = 2\delta(0) \,. \label{infinite}                   
\end{align}
Note that with this definition,  $V$ is dimensionless.  The delta function appears due to the property of Legendre polynomials, 
$2\delta(1-x) = \sum_{\ell=0}^\infty (2\ell+1)P_\ell(x)$, with $P_\ell(0)=1.$ 
The factor in the right hand side of Eq.~\eqref{trA1rhon2}, $\delta(0)/(2\pi k^2 \delta^{(3)}(0))$, 
is equal to $2/V$ (see Eq.~\eqref{threeDdelta}). 
Therefore Eq.~\eqref{trA1rhon2} may be  rewritten as 
\begin{align}
\tr_{A_1}(\rho_{A_1})^n = \biggl({2 \over V}\biggr)^{n-1} \int_{-1}^1 d\!\cos\theta\, (\CP_{11}(\cos\theta))^n \,. \label{trrhoeln}
\end{align}

By using the cross sections \eqref{diffcross}, \eqref{ijcross} and \eqref{totcross}, 
we can rewrite Eq.~\eqref{elp1} as 
\begin{align}
\CP_{11}(\cos\theta) = \delta(1-\cos\theta)\biggl\{ 1 - {\sigma_{11} \over {\pi \over k^2}V -(\sigma_{11}^{\rm tot} -\sigma_{11})}\biggr\}
	+\ {2k^2 \over {\pi \over k^2}V -(\sigma_{11}^{\rm tot} -\sigma_{11})}\ {d\sigma_{11} \over dt} \,. \label{P1crosec}
\end{align}
Note that $\sigma_{11}^{\rm tot} -\sigma_{11}$ is equal to 
the sum of all inelastic cross sections, $\sigma_{21}+\sigma_{11}^X$. 
$\CP_{11}$ depends on the infinite volume of Hilbert space, $V$. 
Refs.\cite{PS1,PS2} have suggested to regularize it by ``volume regularization'', see the next subsection. 

Finally, we obtain the entanglement entropy of the elastic channel, 
\begin{align}
S_{11} = -\lim_{n\to 1}{\partial \over \partial n} \tr_{A_1}(\rho_{A_1})^n 
= \ln{V \over 2} - \int_{-1}^1 d\!\cos\theta\, \CP_{11}(\cos\theta) \ln \CP_{11}(\cos\theta) \,, \label{EEel}
\end{align}
where we used $\int_{-1}^1 d\!\cos\theta\, \CP_{11}(\cos\theta) = 1$.

\subsubsection{Inelastic two-particle channel}

In the center-of-mass frame, the reduced density matrix \eqref{INELredDMref} becomes 
\begin{align}
\rho_{A_2} 
&= {1 \over \CN_2} \int {d^3\vp \over 2E_{A_2\vp}}\,
{\delta(p-k') \delta(0) \over p\sqrt{s}} 
|\bbra{\vp}{2}|{\bf t}|\kket{\vk}{1}|^2\,
|\ket{\vp}{A_2}\,\bra{\vp}{A_2}| \,, \label{rho2}
\end{align}
where 
\begin{align}
k' = \sqrt{(m_{A_2}^2+m_{B_2}^2 -s)^2 -4m_{A_2}^2 m_{B_2}^2 \over 4s} \,.
\end{align}
The details of this calculation are shown in Appendix \ref{sec:deltafunc}.
Note that $k'$ satisfies $E_{A_2k'}+E_{B_2k'} = \sqrt{s}$.
The normalization factor $\CN_2$ is determined by $1 = \tr_{A_2}\tr_{B_2}\rho_2 = \tr_{A_2} \rho_{A_2} $, 
so that 
\begin{align}
\CN_2 = \delta(0) \delta^{(3)}(0) \int d^3\vp\, {\delta(p-k') \over p\sqrt{s}} 
|\bbra{\vp}{2}|{\bf t}|\kket{\vk}{1}|^2 \,. \label{norm2a}
\end{align}
From Eq.~\eqref{rho2}, we calculate 
\begin{align}
\tr_{A_2}(\rho_{A_2})^n 
= \int d^3\vp\, \delta^{(3)}(0) \biggl( {\delta(p-k') \delta(0) \over \CN_2\cdot p\sqrt{s}} 
|\bbra{\vp}{2}|{\bf t}|\kket{\vk}{1}|^2 \biggr)^n \,. \label{tr2rhona}
\end{align}
Substituting the partial wave expansions \eqref{tmatPWE} and \eqref{overlapPWE} 
into Eqs.~\eqref{norm2a} and \eqref{tr2rhona}, 
we obtain
\begin{align}
\tr_{A_2}(\rho_{A_2})^n 
	&= \int d\!\cos\theta\, \biggl({\delta(0) \over 2\pi k'^2 \delta^{(3)}(0)}\biggr)^{n-1} 
(\CP_{21}(\cos\theta))^n \,, \label{trA2rhon}\\ 
\CP_{21}(\cos\theta) &\equiv {|\sum_{\ell=0}^\infty (2\ell+1)\tau_{21}^\ell P_\ell(\cos\theta)|^2 \over 2 \sum_{\ell=0}^\infty (2\ell+1)|\tau_{21}^\ell |^2} \,. \label{p2PW}
\end{align}
By using the cross sections \eqref{diffcross} and \eqref{ijcross}, 
Eq.~\eqref{p2PW} is described as  
\begin{align}
\CP_{21}(\cos \theta) = {2k^2 \over \sigma_{21}}{d\sigma_{21} \over dt} \,. \label{P2crosec}
\end{align}
$\CP_{21}$ satisfies $\int_{-1}^1 d\!\cos\theta\, \CP_{21}(\cos\theta) = 1$
in the same way as $\CP_{11}$. 
Eq.~\eqref{P2crosec} has simpler expression than Eq.~\eqref{P1crosec}, 
because the unit matrix part of S-matrix, ${\bf 1}$, in ${\bf s} = {\bf 1} +2i{\bf t}$ does not 
contribute to $\CP_{21}$ in the inelastic channel, but it does to $\CP_{11}$ in the elastic channel. 
Therefore $\CP_{21}$ is finite and does not depend on the infinite volume of Hilbert space, $V$.

In terms of Eq.~\eqref{threeDdelta}, Eq.~\eqref{trA2rhon} becomes 
\begin{align}
\tr_{A_2}(\rho_{A_2})^n = \biggl({2 \over V}\biggr)^{n-1} \int_{-1}^1 d\!\cos\theta\, (\CP_{21}(\cos\theta))^n \,.
\end{align}
We thus obtain the entanglement entropy, 
\begin{align}
S_{21} = -\lim_{n\to 1}{\partial \over \partial n} \tr_{A_2}(\rho_{A_2})^n 
= \ln{V \over 2} - \int_{-1}^1 d\!\cos\theta\, \CP_{21}(\cos\theta) \ln \CP_{21}(\cos\theta)
\end{align}
By using the Mandelstam variable $t$, 
we can also rewrite $S_{21}$ as 
\begin{align}
S_{21} = \ln{V \over 2} 
	- \int_{-4k^2}^0\!\! \!\!dt\ \ { 1 \over \sigma_{21}}{d\sigma_{21} \over dt} 
\ \ \ln {\biggl({2k^2  \over  \sigma_{21}}{d\sigma_{21} \over dt}\biggl)} \,. 
\label{EEinel}
\end{align}
We note that $S_{11}$ for the elastic two-particle final state 
and $S_{21}$ for the inelastic two-particle final state have a common diverging factor, $\ln(V/2)$.

\subsection{Regularization of infinite Hilbert space volume}

Let us regularize the infinite volume $V$ in the same way as Ref.~\cite{PS2}, which is the volume regularization. 
In Eq.~\eqref{P1crosec}, the first term can be identified as the contribution of non-interacting process. 
Indeed it is generally infinite at $\theta =0$ being proportional 
to the infinite total volume $2 \delta(0)$ of the Hilbert space (see Eq.\eqref{infinite}).
Eq.~\eqref{P1crosec} gives a finite solution only if the proportional factor,  
\begin{align}
1 - {\sigma_{11} \over {\pi \over k^2}{\tilde V} -(\sigma_{11}^{\rm tot} -\sigma_{11})} \,, \label{coeff}
\end{align}
vanishes with the ``regularized volume'' ${\tilde V}$, {\it i.e.}, 
\begin{align}
{\tilde V} = {k^2 \over \pi} \sigma_{11}^{\rm tot} \,. \label{volume}
\end{align}
Hence we are led to set the ``regularized volume'' ${\tilde V}$ dimensionless again. 
It is important to note that  ${\tilde V}$ depends only on the total cross section 
of the incoming particles. 
Therefore, in particular,  ${\tilde V}$ is identical for elastic and inelastic two-body scatterings from the same initial particles.

Then the entanglement entropy for elastic channel \eqref{EEel} is regularized as
\begin{align}
{\tilde S}_{11} 
&= \ln {k^2 \sigma_{11}^{\rm tot} \over 2\pi} 
- \int_{-1}^1 d\!\cos\theta\,{\tilde \CP}_{11} \ln {\tilde \CP}_{11} \,, \quad 
{\tilde \CP}_{11} = {1 \over \sigma_{11}} {d\sigma_{11} \over d\!\cos\theta}\,, \label{vregEEel}
\end{align}
while the entanglement entropy for inelastic channel \eqref{EEinel} is 
\begin{align}
{\tilde S}_{21} 
&= \ln {k^2 \sigma_{11}^{\rm tot} \over 2\pi} 
- \int_{-1}^1 d\!\cos\theta\,\CP_{21} \ln \CP_{21} \,, \quad 
\CP_{21} = {1 \over \sigma_{21}} {d\sigma_{21} \over d\!\cos\theta}\,. \label{vregEEinel}
\end{align}
Note that ${\tilde \CP}_{11}$ is regularized, while $\CP_{21}$ has not to be.
However, it is interesting that ${\tilde \CP}_{11}$ and $\CP_{21}$ finally have identical forms in terms of normalized two-body cross sections.

About the regularization of the elastic channel, one notes that  
if we neglect the non-interacting part in the beginning of our derivation, that is to say, 
if we replace the S-matrix element $\bra{\vp,\vq}{1}| (1\!+\!2i\CT) | \ket{\vk,\vl}{1}$ in Eq.~\eqref{finstel} with $\bra{\vp,\vq}{1}| (2i\CT) | \ket{\vk,\vl}{1}$, 
$\CP_{11}(\cos \theta)$ in Eq.~\eqref{trrhoeln} becomes same as ${\tilde \CP}_{11}$ in Eqs.~\eqref{vregEEel}. 
However the factor $(2/V)^{n-1}$ in Eq.~\eqref{trrhoeln} still remains, 
because it originates from the infinite Hilbert space volume, $\delta(0) / (2\pi k^2 \delta^{(3)}(0))$, with azimuthal symmetry (see Eq.~\eqref{threeDdelta}), and thus one is led to use again the regularized volume ${\tilde V}$ in Eq.~\eqref{volume}.

We can easily generalize these results to the case of the scattering 
including more channels of two-particle final states, \eqref{scchannels}. 
The entanglement entropy of the final state, $A_i B_i$ ($i=2,3,\dots$), 
without regularization is 
\begin{align}
S_{i1} 
&= \ln {V \over 2} 
- \int_{-1}^1 d\!\cos\theta\,\CP_{i1} \ln \CP_{i1} \,, \quad 
\CP_{i1} = {1 \over \sigma_{i1}} {d\sigma_{i1} \over d\!\cos\theta}\,. \label{vregEESi1}
\end{align}
Then let us focus on the discrepancy between the entanglement entropies of final states for a given initial two-body state, 
$A_i B_i$ and $A_j B_j$:
\begin{align}
\Delta S_{i,j} \equiv S_{i1} - S_{j1} 
= - \int_{-1}^1 d\!\cos\theta\, (\CP_{i1} \ln \CP_{i1} - \CP_{j1} \ln \CP_{j1}) \quad 
(i,j=2,3,\dots) \,. \label{deltaSij}
\end{align}
This quantity is independent of regularization, 
because $V$ does not contribute to $\CP_{i1}$ and $\CP_{j1}$. 

Even though the regularization is necessary for the entanglement entropy of elastic channel, 
it is intriguing to consider the difference between the volume-regularized entanglement entropy of 
the elastic final state \eqref{vregEEel} and the one of the inelastic final state \eqref{vregEEinel}:
\begin{align}
\Delta {\tilde S}_{2,1} \equiv {\tilde S}_{21} - {\tilde S}_{11} 
= - \int_{-1}^1 d\!\cos\theta\, (\CP_{21} \ln \CP_{21} - {\tilde \CP}_{11} \ln {\tilde \CP}_{11}) \,. \label{deltaS21}
\end{align}
Eq.~\eqref{deltaSij} and Eq.~\eqref{deltaS21} have formally same expression, which consists 
of cross sections and differential cross sections, in the form of the density distributions generated by ${\tilde \CP}_{11} $ and $ \CP_{21}. $

\subsection{Regularized R{\' e}nyi entropy}

It is useful to notice that, within the same scheme, 
one can formulate the regularized R{\' e}nyi entropy in both elastic and inelastic scattering channels,  {\it i.e.},
\begin{align}
{1 \over 1-n}\ln\tr_{A_i}(\rho_{A_i})^n 
&=\ \ln {{\tilde V} \over 2}
 +{1 \over 1-n}\ln \int_{-1}^1 d\!\cos\theta \,
\bigl({\tilde P}_{i1} (\cos\theta ) \bigr)^n \nonumber\\
&=\ \ln \biggl({k^2 \sigma_{11}^{\rm tot}\over 2\pi} \biggr) +{1 \over 1-n}
\ln \int_{-1}^1 d\!\cos\theta \,\biggl({1 \over \sigma_{i1}}{d\sigma_{i1} \over d\!\cos\theta }\biggr)^n \quad (i=1,2) \,.
\label{renyi}
\end{align}
About this result, we note a difference with Ref.~\cite{LY} for which 
the elastic R{\' e}nyi entropy is proportional to the elastic cross section 
(see, {\it e.g.,} Eqs.~(99) in Ref.~\cite{LY}). 
This discrepancy might be caused by the regularization. 
In our case, the control of the regularization has been done in detail in Ref.\cite{PS1}. 
We found that we may bound the regularized volume \eqref{volume} from below and above 
by regularizing directly partial waves by physical cut-offs \cite{PS2}. 
A more detailed comparison would require a thorough study of the regularization chosen by Ref.~\cite{LY}. It is beyond the scope of our study but may be useful in the future.

\section{Evaluation of entanglement entropy}
\subsection{Elastic and inelastic cross sections in the two-nucleon sector}

We shall evaluate entanglement entropies in a high-energy proton-neutron scattering, 
as a prototype for the comparison of elastic and inelastic two-body cases. 
Their quantitative estimates can be performed using the neutron-proton scattering data. More precisely,  the inelastic neutron-proton charge exchange reaction $pn\to np$ has been measured with a high-momentum  neutron beam  on a proton target \cite{KDLO}. The quantitative description of the elastic $pn\to pn$ cross sections can be obtained from the same two-body scattering  with a forward neutron. One also may use data from the high-energy  $pp\to pp$ scattering in the same energy  range whose cross sections  are equivalent, at least in the low momentum transfer range \cite{De}, as well as total cross sections. We thus are able to make use of the well documented phenomenology in that elastic case \cite{Kwak}.

\begin{figure}[h]
\centering
\includegraphics{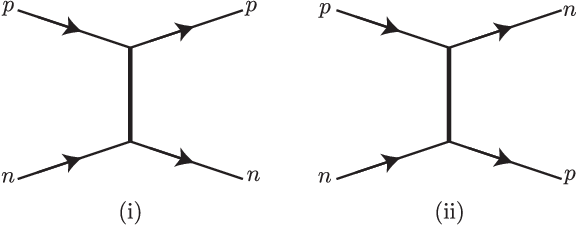}
\caption{(i) the elastic channel: $pn\to pn$. (ii) the inelastic channel: $pn\to np$.}\label{pnscattering}
\end{figure}

We thus focus on the following two-particle channels (see Fig.~\ref{pnscattering}): 
\begin{numcases}
{pn \to}
	pn & (elastic channel) \\
	np & (inelastic channel) 
	\label{channels}
\end{numcases}
where the first of the final particles is in the near forward (small $ t$) range.
We can recognize $pn \to np$ as an inelastic channel with charge one current exchange. 
In order to concretely evaluate the entanglement entropies, \eqref{vregEEel} and \eqref{vregEEinel}, 
the differential elastic and inelastic cross sections as functions of $t$ and the total $pn$ cross section are necessary. 

For our analysis, we take advantage of obtaining the near forward proton-neutron elastic differential cross section 
from a precise phenomenological description of related amplitudes obtained from data on proton-proton elastic scattering, 
in Ref.~\cite{PB} for differential cross sections in the forward region and in  Refs.~\cite{BC1,BC2} for total cross sections. 
The inelastic differential cross section in the forward region is provided by Refs.~\cite{BouD,BizD}.
One finds the parameterizations of those observables used for our analysis in Appendix \ref{sec:param}.

\subsection{Evaluation of the entanglement entropies in the two-nucleon sector}

By using those fitting functions shown in the previous subsection, 
let us evaluate the elastic and inelastic entanglement entropies. 
Using the change of variable, $\cos\theta \to t$, by means of the relation \eqref{Mandelstam}, one writes the corresponding relation between density distributions, 
\begin{align}
{\tilde \CP}(\cos\theta)\, d\!\cos\theta = {P}(t)\, dt\,.
\end{align}
One then infers from Eqs.\eqref{vregEEel} and \eqref{vregEEinel}, and the notations therein, 
that $P_{\rm el}(t)$ ({\it resp}.~$P_{\rm inel}(t)$) defined 
by $P_{\rm el}(t) = {2 \over {s-4m^2}}\ {\tilde \CP}_{11}(\cos\theta)$ 
({\it resp}.~$ P_{\rm inel} (t) = {2 \over s-4m^2} \,{\tilde \CP}_{21}(\cos\theta)$), 
where $m$ is the nucleon mass, are the density distributions of the  elastic ({\it resp}.~inelastic) collisions in the transfer variable $t.$
 Naturally,  $P_{\rm el, inel}(t)$ satisfy $\int dt\, P_{\rm el, inel}(t) = 1$. 

We then rewrite the regularized entanglement entropies, 
\eqref{vregEEel} and \eqref{vregEEinel}, as 
\begin{align}
{\tilde S}_{\rm el} &= - \int_{-4k^2}^0 dt\, P_{\rm el}(t)\, \ln {\left( {4\pi \over \sigma_{\rm tot}} P_{\rm el}(t)\right)} \,, \label{formelEE}\\
{\tilde S}_{\rm inel} &= - \int_{-4k^2}^0 dt\, P_{\rm inel}(t)\, \ln {\left( {4\pi \over \sigma_{\rm tot}} P_{\rm inel}(t)\right)} \,. \label{forminelEE}
\end{align}
It is important to note that the lower limit of integration  in these equations is given by pure kinematics. But the physical limitation to the parameterizations for both elastic and inelastic cases are concentrated in their respective forward region, {\it i.e.}~at small momentum transfer region.
For instance, in the $pn$ sector, $ P_{\rm el}(t)$ and $ P_{\rm inel}(t)$ are governed by very different parametric functions decreasing fastly with momentum transfer.

We numerically compute the elastic entanglement entropy \eqref{formelEE} 
in terms of \eqref{totCS}, \eqref{pnpnDifCS} and \eqref{pnpnCS}.
${\tilde S}_{\rm el}$ has a local minimum at $\sqrt{s}=41.7$ GeV.
We then compute the inelastic entanglement entropy \eqref{forminelEE} 
in terms of \eqref{totCS}, \eqref{pnnpDifCS} and \eqref{pnnpCS}.
${\tilde S}_{\rm inel}$ has a local minimum at $\sqrt{s}=12.4$ GeV.
\begin{figure}[h]
\centering
\includegraphics{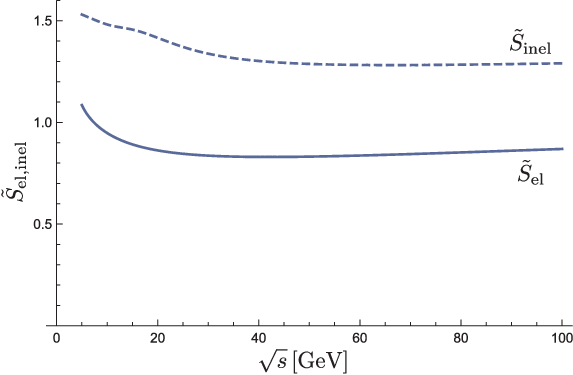}
\caption{The elastic entanglement entropy ${\tilde S}_{\rm el}$ (solid line) and the inelastic one ${\tilde S}_{\rm inel}$ (dashed line).}\label{pnvsnp}
\end{figure}
Both entropies are depicted  in Fig.~\ref{pnvsnp} as functions of the center-of-mass energy $\sqrt{s}.$
The inelastic entanglement entropy is larger than the elastic one as shown in Fig.~\ref{pnvsnp}.

\begin{figure}[h]
\centering
\includegraphics{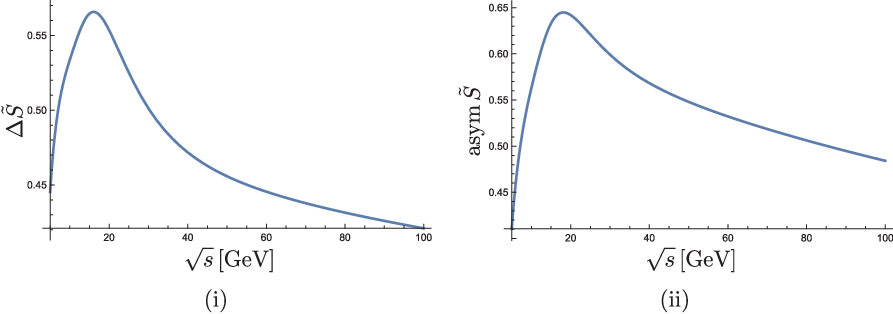}
\caption{(i) The difference $\Delta {\tilde S}$ and (ii) the asymmetry ${\rm asym}\,{\tilde S}$ of entanglement entropies between the elastic and inelastic channels.}\label{diffasym}
\end{figure}
We also show in Fig.~\ref{diffasym} the discrepancy \eqref{deltaS21} and the asymmetry between these two entropies, namely, 
\begin{align}
\Delta {\tilde S} = {\tilde S}_{\rm inel} - {\tilde S}_{\rm el} \,, \quad 
{\rm asym}\,{\tilde S} = {{\tilde S}_{\rm inel} - {\tilde S}_{\rm el} \over {\tilde S}_{\rm el}}. 
\end{align}
One can observe that at lower energy the elastic entanglement entropy decreases faster than the inelastic one while it is the contrary at higher energy 
as shown in Fig.~\ref{diffasym}.

The difference observed between the elastic and inelastic entanglement entropies 
for identical initial nucleons can be qualitatively interpreted 
using a simple exponential approximation of the dominant cross sections, 
following the treatment in Refs.~\cite{PS1,PS2}.                         
Let us write 
\begin{align}
{d\sigma_{\rm el} \over dt} \sim \sigma_{\rm el}(s) \,e^{R^2_{\rm el}t} \,, \quad
{d\sigma_{\rm inel} \over dt} \sim \sigma_{\rm inel}(s) \, e^{R^2_{\rm inel}t} \,.
\label{sigma}
\end{align}
$R_{\rm el}$ and $R_{\rm inel}$ are the effective radii. 
Using formulas \eqref{formelEE} and \eqref{forminelEE}, one easily finds
\begin{align}
&{\tilde S}_{\rm el} = \ln {\left( {\sigma_{\rm tot} \over 4\pi R^2_{\rm el}}\right)} + 1 \,, \\
&{\tilde S}_{\rm inel} = \ln {\left( {\sigma_{\rm tot} \over 4\pi R^2_{\rm inel}}\right)} +1 \,,\\
&\Delta {\tilde S} = {\tilde S}_{\rm inel} - {\tilde S}_{\rm el} = \ln {\left( {R^2_{\rm el}}\over  {R^2_{\rm inel}}\right)} \,.
 \label{Entropy}
\end{align}
Hence, the inelastic scattering being smoother in momentum transfer than the elastic one at comparable energies, see Refs.~\cite{KDLO,Kwak}, one has $ R^2_{\rm inel} < R^2_{\rm el}$ and thus  $ S_{\rm inel} > S_{\rm el},$ as  quantitatively confirmed by the full calculations, see Fig.~\ref{pnvsnp}. Smaller is the effective radius, larger is the entanglement between the final couple of particles.

\subsection{Density of entanglement entropy in transverse momentum}

The integral equations for entanglement entropies \eqref{formelEE} and \eqref{forminelEE}
can be rewritten
\begin{align}
{\tilde S}_{\rm el} &= \int_0^{4k^2} d|t|\,D_{\rm el}(t) \,, \quad 
D_{\rm el}(t) \equiv -P_{\rm el}(t)\, \ln {\left( {4\pi \over \sigma_{\rm tot}} P_{\rm el}(t)\right)} \,,\label{dSformelEE}\\
{\tilde S}_{\rm inel} &= \int_0^{4k^2} d|t|\,D_{\rm inel}(t) \,, \quad 
D_{\rm inel}(t) \equiv -P_{\rm inel}(t)\, \ln {\left( {4\pi \over \sigma_{\rm tot}} P_{\rm inel}(t)\right)}  \,. \label{dSforminelEE}
\end{align}
Note again, as for Eqs.~\eqref{formelEE} and \eqref{forminelEE}, the upper bound of intgration is given by kinematics, but the physical focus (together with the parameterizations) is on the different forward regions of the two reactions.

Thus we obtain the expressions for the entropy densities, $D_{\rm el}$ and $D_{\rm inel}$, 
as functions of the transfer momentum $q_T=\sqrt{|t|}$.
Note that $D_{\rm el,\rm inel} $ as well as $P_{\rm el,\rm inel} $ are also functions of the energy variable which is implicit.

\begin{figure}
\centering
\includegraphics{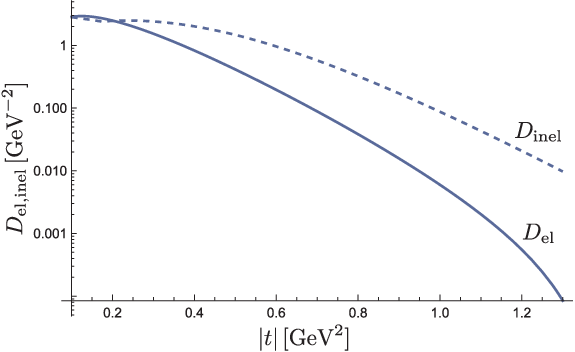}
\caption{The elastic  density  $D_{\rm el}$ (solid line) and the inelastic one $D_{\rm inel}$ (dashed line) as functions of $|t|$ at $\sqrt{s}=20$ GeV}\label{dSdt_el-inelB}
\end{figure}
\begin{figure}
\centering
\includegraphics{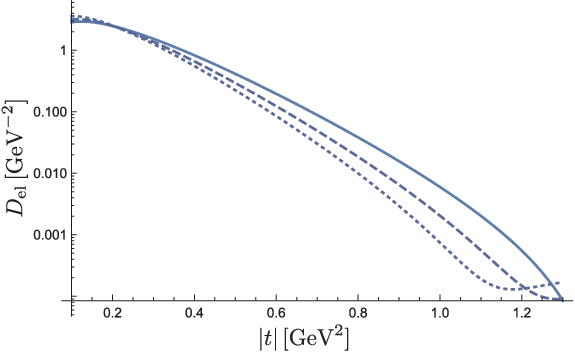}
\caption{The elastic density $D_{\rm el}$ as a function of $|t|$ 
	at $\sqrt{s}=20$\,GeV (solid line), 50\,GeV (dashed line) and 100\,GeV (dotted line).}
\label{eld}
\end{figure}
\begin{figure}
\centering
\includegraphics{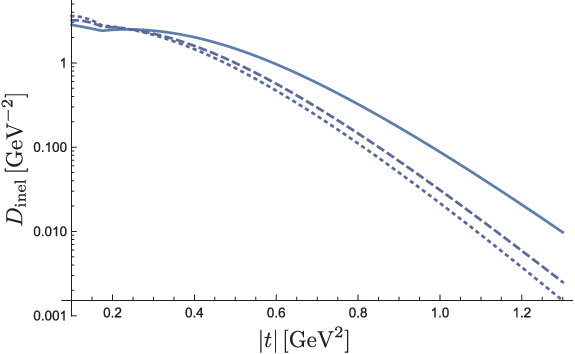}
\caption{The inelastic density $D_{\rm inel}$ as a function of $|t|$ at $\sqrt{s}=20$\,GeV (solid line), 50\,GeV (dashed line) and 100\,GeV (dotted line).}
\label{ineld}
\end{figure}
Indeed, the physical interpretation of these densities are interesting: they describe 
the local amount of entanglement entropy flow as a function of momentum transfer for 
a given total energy $\sqrt{s}$. As depicted in 
Fig.~\ref{dSdt_el-inelB}, the elastic entropy flow $D_{\rm el}$ 
and the inelastic one $D_{\rm inel}$ at $ \sqrt{s}=20$ GeV are both stronger and similar in the very forward direction. They both decrease when the  momentum transfer increases but the  inelastic one then dominates by orders of magnitude. This is the cause of the different  trend revealed in  Fig.~\ref{pnvsnp} for the overall entanglement entropies.

One thus understands that entanglement is stronger when the particle trajectories are very near-by as expected but the exchange of a non-vacuum quantum number, namely the electric charge in our case, seems to keep  on longer distances the entanglement between the final particles. This property obtained in the two nucleon sector would be interesting to be investigated elsewhere.

Interestingly, the quasi equality between both densities in the very forward region seen in Fig.~\ref{dSdt_el-inelB} seems to indicate that the selection process for a proton or a neutron  in the very forward scattering, which has a strong consequence for the hierarchy of  the energy dependence of the differential cross sections, does not  modify the entanglement in this very forward region. By contrast, both for the elastic or the inelastic 
scattering considered as a function of their own energy dependence, 
there is a similar trend of the densities as shown by Figs.~\ref{eld} and \ref{ineld} despite the very different energy dependence of the differential cross sections.

\section{Conclusion and outlook}
Let us summarize our results. 

We have performed the formalism allowing to derive the entanglement entropies 
in parallel for elastic and inelastic two-body reactions at high energy. 
Once using the``volume'' regularization characterized by the total cross sections of the incident particles, 
the elastic and inelastic entanglement entropies have a similar expression, 
and enable us to study various relations between the entanglement entropy for different final two-body state channels. 

As a typical application of the formalism and an example of the mentioned relations, we have evaluated the entanglement entropy in the two-nucleon sector, namely, 
$pn\to pn,$ in the forward region for the elastic scattering, 
and $pn\to np$ in the forward region for the inelastic scattering. 

Remarkably, using parameterizations of data in the same energy range 
at $\sqrt{s}=$20, 50 and 100 GeV for both elastic and inelastic $pn$ scatterings, 
we find a marked difference between their entanglement entropies, 
the inelastic entanglement entropy  being  significantly stronger than the elastic one. 
By contrast, the energy dependence in both channels is rather mild, 
with a slight increase at higher energies. 
This is contrasted with the cross-section dependence which are polynomially decreasing  for the inelastic channels contrary to the elastic one.

As an interesting output of our formulation, we obtain the density of entanglement entropy 
as a function of transverse momentum squared $q_T^2 = |t|$, 
which allows for a detailed description of the entropy flow in the various reactions. 
Applied to the two-nucleon sector, we find that both densities of the elastic and inelastic cases are equal in the very forward region $|t|$, 
while the inelastic density more and more dominates the elastic one, explaining the difference in the overall entropy. 

In outlook, we would like to emphasize some interesting further studies allowed by our formalism. 
At first, it will be interesting to evaluate the elastic-inelastic channel relation for all cases where data exist. This would help answering quite fundamental questions:
\begin{itemize}
\item 
Is there a common feature that vacuum quantum number exchange (elastic case) would generate less entanglement than non-vacuum exchange (inelastic case)?
\item 
The study of the entanglement entropy density as a function of transverse momentum seems to open a window on the flow of entanglement in a two-body scattering. Is there some regularity of this flow for different initial particles indicating some universality properties?
\end{itemize}
All in all, the goal is to understand the key features of  entanglement  in momentum due to the interaction and understand which are the order parameters to which it obeys: initial particle quantum numbers, exchange quantum numbers, eventually some particular energy ranges, {\it etc.}.

\bigbreak\bigskip\bigskip
\centerline{{\bf Acknowledgments}}\nobreak

S.~S.~was supported in part by JSPS Grant-in-Aid for Scientific Research (C) \#22K03625, 
and is grateful to Institut de Physique Th{\' e}orique (IPhT), CEA-Saclay for their hospitality.
\bigbreak\bigskip\bigskip

\begin{appendix}

\section{Dirac delta function and Legendre polynomials}

The three-dimensional Dirac delta function in spherical coordinates with azimuthal symmetry can be written as 
\begin{align}
\delta^{(3)}(\vp-\vk) = {\delta(p-k) \over 4\pi k^2} \sum_{\ell=0}^\infty (2\ell+1) P_\ell(\cos\theta) \,,
\end{align}
with $\cos\theta = (\vp\cdot\vk)/(pk)$. 
Taking the limit $\vp \to \vk$, we obtain 
\begin{align}
{4\pi k^2 \delta^{(3)}(0) \over \delta(0)} = \sum_{\ell=0}^\infty (2\ell+1) P_\ell(1)
= \sum_{\ell=0}^\infty (2\ell+1) \,. \label{threeDdelta}
\end{align}

\section{Dirac delta function for energy conservation}\label{sec:deltafunc}

We fix the center-of-mass energy of two incoming particles as 
\begin{align}
\sqrt{s} = E_{A_j\vk}+E_{B_j\vk} =\sqrt{m_{A_j}^2 +k^2} +\sqrt{m_{B_j}^2 +k^2}\,.
\end{align}
Then we compute the delta function for energy conservation, 
\begin{align}
&\delta((E_{A_h\vq}+E_{B_h\vq})-(E_{A_j\vk}+E_{B_j\vk})) = \delta((E_{A_h\vq}+E_{B_h\vq})-\sqrt{s}) \nonumber\\
&= \delta\biggl({q\sqrt{s} \over E_{A_h\vq}E_{B_h\vq}}(q-q') + \CO((q-q')^2) \biggr) \nonumber\\
&= {2E_{A_h\vq}\, 2E_{B_h\vq} \over 4q\sqrt{s}} \delta(q-q') \,. \label{eneconvdeltaI}
\end{align}
$q'$ is given by $(E_{A_h\vq'}+E_{B_h\vq'})-\sqrt{s} = 0$, that is, 
\begin{align}
q' = \sqrt{(m_{A_h}^2+m_{B_h}^2 -s)^2 -4m_{A_h}^2 m_{B_h}^2 \over 4s} \,.
\end{align}

\section{Parameterizations of the elastic and inelastic $pn$ cross sections.} \label{sec:param}

\subsection{Total cross section} \label{sec:paramtot}

The same $pn$ total cross section appears in both the elastic and inelastic entanglement entropies, 
\eqref{vregEEel} and \eqref{vregEEinel}. 
We can read it from the parameterization of the total $pp$ cross sections, arguing from the equivalence with $pn$ forward scattering. From Refs.\cite{BC1,BC2}, one reads 
\begin{align}
\sigma_{\rm tot} &= -{1 \over s}\Im (M_+(s) - M_-(s)) \ [{\rm GeV}^{-2}]\,, \label{totCS}
\end{align}
with
\begin{align}
\Im M_+(s) = -s \bigl(163 -20.4 \ln s +1.75 (\ln s)^2 \bigr) \,, \quad
\Im M_-(s) = -73.1 s^{0.48} \,.
\label{Regge}
\end{align}
By definition in \cite{BC1,BC2}, $M_\pm$ imply amplitudes such that $M_\pm = {1 \over 2}(M_{pp}\pm M_{p{\bar p}}).$

\subsection{Elastic cross sections} \label{sec:paramel}

In Ref.~\cite{PB} one can read the elastic $pp$ differential cross section (equivalent to   $pn\to pn$ ones in the forward region);
\begin{align}
{d\sigma_{\rm el} \over dt} = \biggl| i\, 6.88 e^{\left(2.69+0.306\ln s -i\,0.153\pi\right)t} +\biggl({10.3 \over s}-i\,0.035 \biggr)e^{0.89 t}\biggr|^2 \ [{\rm mb}\cdot{\rm GeV}^{-2}]\,. \label{pnpnDifCS}
\end{align}
Then the elastic cross section is given, neglecting the exponentially small tail at large transfer, by 
\begin{align}
\sigma_{\rm el} = \int_{-\infty}^0 dt\, {d\sigma_{\rm el} \over dt} \,. \label{pnpnCS}
\end{align}

\subsection{Inelastic cross sections} \label{sec:paraminel}

Refs.~\cite{BouD,BizD} have shown three models of the inelastic differential cross section 
for $pn \to np$ with similar overall results.
We adopt the two-component model \cite{BouD} using the first component defined in 
 \cite{BizD}, which is based on the five helicity amplitudes. 

The inelastic differential cross section based on the five helicity amplitudes \cite{BouD,BizD} is  
\begin{align}
{d\sigma_{\rm inel} \over dt} = {2\pi \over s^2} 0.389 \left(|\varphi_1|^2+|\varphi_2|^2+|\varphi_3|^2+|\varphi_4|^2+|\varphi_5|^2\right) \,. \label{pnnpDifCS}
\end{align}
An important notice: since we are interested here in the entanglement on the momentum Hilbert space  only, it is well known that one has to make the overall trace over the
 spin or  helicity degrees of freedom. This  results in the summation in Eq.\eqref{pnnpDifCS}. 
 Would it be a study of the entanglement in
 spin or  helicity degrees of freedom joined or not with the momentum one, one would have to make  partial traces also on these degrees of freedom.
 
In practice, 
$\varphi_i$ is split into an energy-independent part $\varphi_i^c$ and an energy-dependent part $\varphi_i^v$, {\it i.e.}, 
$\varphi_i = \varphi_i^c+\varphi_i^v$. 
From Refs.~\cite{BouD,BizD} we write down $\varphi_i^c$, 
\begin{align}
\varphi_1^c &= 15 {t \over s-4m^2}{m^2 \over s}(-2.2)e^{5t} \,, \\
\varphi_2^c &= 15 \left\{{t \over t-m_\pi^2} +{4t \over m^2} -0.9 -{0.45 t \over s-4m^2} +{0.65t \over s-4m^2}{m^2 \over s} +{0.65 t \over s-4m^2}\left({m^2 \over s}+0.5\right)\right\}e^{5t} \,, \\
\varphi_3^c &= 0 \,, \\
\varphi_4^c &= 15 \left\{ {t \over t-m_\pi^2} +{4t \over m^2} +{t \over s-4m^2}\left(1.55{m^2 \over s}+0.325\right) \right\}e^{5t} \,, \\
\varphi_5^c &= 15 {m^2 \over s-4m^2} \sqrt{-{t \over 4m^2}{s-4m^2+t \over s}}2.2e^{5t} \,, 
\end{align}
where the nucleon mass, $m=0.939$\,GeV, and the pion mass, $m_\pi=0.139$\,GeV. 
One can reads $\varphi_i^v$ from Eqs.~(2a,b,c) and Eqs.~(4b,c) in \cite{BouD} as follows:
\begin{align}
&\varphi_1^v = \varphi_3^v = \beta_{\rm eff}(t) s^{\alpha_{\rm eff}(t)} \,, \label{phiv1} \\
&\varphi_2^v =-\varphi_4^v = 17t \beta_{\rm eff}(t) s^{\alpha_{\rm eff}(t)} \,, \label{phiv2}\\
&\varphi_5^v = -\sqrt{-17t}\beta_{\rm eff}(t) s^{\alpha_{\rm eff}(t)}\,,\label{phiv3}
\end{align}
where 
\begin{align}
\alpha_{\rm eff}(t) &= 
	\begin{cases}
	0.74+0.67t &(|t| < 0.175)\\
	0.63+0.04t &(|t| > 0.175)
	\end{cases} \\
\beta_{\rm eff} &= -0.12 e^{5.5t} \,.
\end{align}
Then the inelastic cross section is calculated through 
\begin{align}
\sigma_{\rm inel} = \int_{-\infty}^0 dt\, {d\sigma_{\rm inel} \over dt} \,. \label{pnnpCS}
\end{align}

\subsection{Uncertainties} \label{sec:uncertaintes}

The error bars obtained for the total $pn$ cross-sections  and for the differential elastic ones  are negligible, thanks to the use of the equivalent $pp$ total \cite{BC1,BC2} and differential \cite{PB} cross-sections whose data and their phenomenological description are well documented.

For the inelastic $pn\to np$  cross sections the two-component model we considered leads to a $\chi^2/6 \leq 1$ for six energies and for each of 24 among 28 data points in $|t|$ \cite{BouD}. This is quite satisfactory taking into account that experimental results are less precise due to the necessary use of  neutron incident particles.

\end{appendix}

\end{document}